\documentclass[a4paper, 12pt, oneside]{article}
\usepackage[cp1251]{inputenc}
\usepackage{amssymb}
\usepackage{amsmath}
\usepackage{cite}
\usepackage{graphicx}
\usepackage{afterpage}

\makeatletter
\renewcommand{\@biblabel}[1]{#1.\hfill}
\begin{document}

\begin{titlepage}
\title{\center \large \textbf{Conjugation in hydrogen-bonded systems}\\
\vspace{10mm} \large {Yulia V. Novakovskaya}\\
\vspace{15mm} \large \textit{Moscow State University,\\ Leninskie Gory, Moscow, 119991 Russia \\ E-mail: jvnovakovskaya@gmail.com}}
\author{}
\date{}
\maketitle 
\begin{abstract}
Analysis of the electron density distribution in clusters composed of hydrogen fluoride, water, and ammonia molecules, especially within the hydrogen-bond domains, reveals the existence of both $\sigma$- and $\pi$-binding between molecules. The $\sigma$-kind density distribution determines the mutual orientation of molecules. A $\pi$-system may be delocalized conjugated, which provides additional stabilization of molecular clusters. In those clusters where the sequence of hydrogen bonds is not planar, a peculiar kind of $\pi$-conjugation exists. HF$_2^-$ and H$_5$O$_2^+$ ions are characterized by quasi-triple bonds between the electronegative atoms. The most long-lived species stabilized by delocalized $\pi$ binding are rings and open or closed hoops composed of fused rings. It is conjugated $\pi$ system that determines cooperativity phenomenon. 

\vspace{5mm}

\end{abstract}
\thispagestyle{empty}
\end{titlepage}

\section{Introduction}
A fruitful idea abot the existence of a special kind of intermolecular bonds between such particles as water or ammonia was put forward already a century ago~\cite{Moore}. Although T.S. Moore and T.F. Winmill~\cite{Moore} spoke only about a weak union formed by the molecules, and the concept of hydrogen bond developed later, the idea was already born. M.L. Huggins~\cite{Huggins} and W.M. Latimer and W.H. Rodebush~\cite{Latimer} are usually considered as the originators of the notion, but in its conventional variant it appeared and was accepted by the scientific community due to L. Pauling’s~\cite{Pauling1}. Since that time, the scope of information about hydrogen bonds in the most diverse molecular systems and ensembles has kept increasing. As was mentioned in~\cite{IUPAC_tech}, if considering only the three-year period of 2006–2008, there appeared more than 31000 papers in English, devoted to hydrogen bonding indexed in SciFinder. The amount of papers published in all languages for the last three quarters of a century would be impossible to estimate. It might even exceed a quarter of a million. 

It may be that no other problem of structure theory attracted so continuing attention of researchers. The reasons are quite clear. Hydrogen bonds dramatically change the properties of molecular ensembles; and H-bonded ensembles are present nearly everywhere: in biological species, in the atmosphere, in many areas of technology. Plants and all living organisms are composed of water, up to 90–99\%. Tetrahedral arrangement of water molecules determined by the peculiarities of the electron density (charge) distribution was the first principal discovery in the field~\cite{Bernal}. Later on, the scientific community realized that proteins exist in the form of $\alpha$-helices and $\beta$-sheets~\cite{Pauling}, DNA molecules form double helices~\cite{Watson}, and it is hydrogen bonds that stabilize these structures. Thus, all biological processes proceed in H-bonded environments and involve molecular structures with H-bonds. Insofar as the majority of technological processes proceed in aqueous solutions, and, which is more important, modern technology attempts to mimic the nature on the micro- and nanoscales, we can say that most of the technological schemes are also determined by peculiar properties of H-bonded substances. Finally, the formation of mist and clouds and the relevant processes in the atmosphere, both the troposphere and the stratosphere, are also determined by the ability of water molecules to aggregate either in the absence or in the presence of foreign particles. \textit{Thus, everything in and around us is related one way or another to the properties of H-bonded systems, but we still have no clear concept of H-bonding.} 

\section{New definition of hydrogen bond: progress and unsolved problem}

Having analyzed the milestone works in the field (refs. in ~\cite{IUPAC_tech}), members of the IUPAC Task Group on Categorizing Hydrogen Bonding and Other Intermolecular Interactions recommended in 2011 the following definition~\cite{IUPAC_def}: {\textquotedblleft}The hydrogen bond is an attractive interaction between a hydrogen atom from a molecule or a molecular fragment X--H in which X is more electronegative than H, and an atom or a group of atoms in the same or a different molecule, in which there is evidence of bond formation.{\textquotedblright} X--H is the bond donor, while Y or Y--Z is the acceptor with an electron rich region, such as a lone pair of Y or a $\pi$-bonded electron pair of Y--Z. 

The definition is accompanied by detailed comments and lists of criteria and characteristics of hydrogen bonds. The main features are summed up below~\cite{IUPAC_def}:
\newcounter{N}
\begin{list}{(\arabic{N})}{\usecounter{N} \labelwidth=4mm \labelsep=2mm \listparindent=0mm}
\item forces that determine the formation of a hydrogen bond have an electrostatic origin, arise from charge transfer between the donor and acceptor, and involve the dispersion component;
\item electron density distribution in the X--H\dots Y--Z domain is usually characterized by a bond path between H and Y with a (3, --1) critical point;
\item H-bond energy correlates well with the extent of charge transfer between the donor and the acceptor; and the strength of H\dots Y bond increases with an increase in the electronegativity of X;
\item X--H\dots Y angle is usually straight, and the closer it to 180$^{\circ}$, the stronger the hydrogen bond and the shorter the H\dots Y distance; accordingly, hydrogen bonds are directional, which affects packing in the corresponding crystals;
\item X--H distance is larger when the fragment is involved in hydrogen bond, and the larger the distance, the stronger the H\dots Y bond; the H-bond formation is accompanied by the decrease in the X--H stretching frequency, the increase in the intensity of the corresponding infrared absorption line, and the appearance of new vibrations;
\item X--H\dots Y--Z hydrogen bond has characteristic NMR signatures including the proton deshielding, spin–spin coupling between X and Y, and nuclear Overhauser enhancements;
\item hydrogen bonds are involved in proton-transfer reactions (X--H\dots Y $\rightarrow$ X\dots H--Y) and may be considered as partially activated precursors of the reactions;
\item networks of hydrogen bonds show the phenomenon of cooperativity that is manifested in deviations of the properties from pair-wise additivity.
\end{list}

The features are not included in the main definition body, because there are hydrogen bonds that possess not all, but only some properties. As examples, one can recall either blue-shifting X--H\dots$\pi$ bonds (X = O, N, C)~\cite{Hobza} or very weak F--H\dots Rg bonds (Rg = Ne, Ar) (which were classified as hydrogen bonds only due to the existence of an H\dots Rg bond path with a (3, --1) critical point)~\cite{Bader}.

Thus, all the characteristics of H-bonds, which were gradually discovered for the three quarters of a century, are accurately systematized now, but we still have no unique concept at out disposal. The concept is strongly needed, because until we clarify the nature of hydrogen bonds, we cannot progress in explaining the aforementioned geometrical and energetic properties of H-bonded systems and their experimental manifestations, especially the cooperativity phenomenon that underlies the functionality of all, primarily biological, species. Being unable to explain, we have no means for founded directional production of systems with desirable structures and properties, which is a serious obstacle on the way of developed nano technology.

\section{Two main kinds of hydrogen bonds}

H-bond energies depend on the nature of interacting molecules, their mutual arrangement, and coordination neighborhood and fall in a broad range, from 0.5 to 40 kcal/mol. The lowest energy is typical of Rg\dots HF complexes, while the highest is observed in H$_5$O$_2^+$ (36 kcal/mol) and HF$_2^-$ (40 kcal/mol) systems, which have symmetric configurations where a proton is located accurately at the midpoint between two oxygen or fluorine nuclei so that the structures are stabilized by two equivalent H\dots X bonds (X = F or O) rather than one covalent and one hydrogen bonds (X--H\dots X). Nuclear configurations of both molecular ions look as if these are unstable intermediates in the symmetric proton transfer reactions H$_3$O$^+$\dots H$_2$O $\rightarrow$ H$_2$O\dots H$_3$O$^+$ or FH\dots F$^-$ $\rightarrow$ F$^-$\dots HF. However, the configurations are stable when there is no external perturbation. When a perturbation appears, they do become unstable intermediates. A reason for that cannot be found if one bases the consideration on the above definition of hydrogen bond.

Then, the following question arises: what is the kind of electron binding that determines the formation of hydrogen bonds? Is it reasonable to consider the interaction as no more than a perturbation of individual molecules, which is implicitly implied when one distinguishes different forces that contribute to the formation of hydrogen bonds? The approach is justified when one deals with typical intermolecular bonds of van-der-Waals kind, that is, when two approaching particles slightly affect the electron density distributions of each other, and the interaction can be estimated as a sum of the first- and second-order perturbation-theory corrections to the zero-order value that corresponds to non-interacting particles. The first-order contribution is the averaged electrostatic energy, while the second-order correction is the polarization (induction and dispersion) energy. Note that for estimating both first- and second-order energy corrections, it is sufficient to know the zero-order functions solely. However, being agglomerated in H-bonded clusters or ensembles, molecules no longer can be treated as independent individual particles, since they acquire novel properties; and \textit{hydrogen bonds formed (in most cases) have more in common with ordinary chemical bonds than with van-der-Waals bonds.} Let us found and develop the idea.

Note that molecules such as hydrogen fluoride, water, and ammonia, as well as their so-to-say derivatives, namely molecules involving OH and NH groups, can form H-bonded networks. At the same time, there are examples of H-bonded molecular pairs that cannot be joined with each other. Therefore, it is reasonable to distinguish \textit{two kinds of hydrogen bonds}: (i) \textit{separate bonds} formed each by two particles and (ii) \textit{bond networks} that unite a large number of molecules. The aforementioned weakest bonds are typically of the former kind, while the strongest bonds, in which FH, OH, or NH groups are involved, are of the latter kind. 

Bond networks appear when each involved particle can form no less than two bonds, one as a proton donor ($d$) and the other as an acceptor ($a$). This is possible when there is a proton, which is chemically bound to an electronegative atom, and a {\textquotedblleft}free{\textquotedblright} valence electron density (uninvolved in interatomic bonds). The closer the proton and the {\textquotedblleft}free{\textquotedblright} density domain in the particle, the stronger the correlation between hydrogen bonds. In other words, when hydrogen bonds are spatially close, i.e., separated with one intramolecular bond (X--H\dots X--H\dots X--H\dots), a system of interacting bonds rather than just a spatial bond network originates. Below, we analyze electronic structures of the clusters composed of water, ammonia, and hydrogen fluoride molecules in order to clarify the character of the electron density distribution that determines the internally consistent state of molecular ensembles.

Insofar as there is a large scope of data about cluster systems composed of water, ammonia, and hydrogen fluoride molecules, we believe it unreasonable and impossible to quote all of works and, at the same time, disrespectful to quote only some of them and neglect the residual. Therefore, the discussion below contains no reference to independent studies. A representative list of references can be found in the aforementioned technical report of the IUPAC Task Group~\cite{IUPAC_tech}.

\section{Electron density distribution: CO LCMO approach}

Undoubtedly, molecular orbital is a conditional mathematical construction. Nevertheless, electrons are physical objects, and spatial distributions of their charges can be described by certain functions. Until we clarify the physical nature of the objects, we can construct only approximate functions that represent (more or less accurately) the mean exterior of the real charge distribution. Then, at present moment, the most reasonable approach is to consider natural orbitals, in the basis of which the total electron density of a molecular system $\rho(\mathbf{r})$ is written as a sum of the weighted densities that correspond to individual $\phi_i(\mathbf{r})$ orbitals:
\begin{equation*}
\rho(\mathbf{r}) = \sum\limits_{i=1}^{P}n_i |\phi_i(\mathbf{r})|^2.
\end{equation*}
When $n_i$ weights (or coefficients), which are referred to as effective occupancies of the orbitals, are close to one or two (i.e., negligibly differ from these integers), one can believe (just believe, since this cannot be proved) that $\phi_i(\mathbf{r})$ functions, which were found as a result of nonempirical calculations, represent reasonable approximations of the actual charge distributions of individual electrons. 

In the case of water, ammonia, and hydrogen fluoride molecules (in the ground electronic states), as well as clusters composed of these molecules, $n_i$ numbers in the above equation meet this requirement; and natural orbitals constructed in the second-order of the Moeller-Plesset perturbation theory (MP2), as well as with the use of fully or partially variation schemes, such as single- or multi-reference configuration interaction method or coupled clusters approach, are very close to the solutions obtained at the Hartree-Fock (HF) level. This means that if we were interested only in the orbital picture of the systems, we could restrict the consideration by the HF level. However, when estimating relative stabilities of diverse cluster configurations or dissociation energies, we cannot neglect the correlation energy contributions. 

Note that it is MO LCAO scheme (where molecular orbitals are approximated by linear combinations of atomic orbitals) that enables one to propose the most straightforward physical interpretation of the quantum chemical description of the electronic subsystems of individual molecules. It is founded when a single-determinant description of the electronic system provides reasonable results. Let us use a similar approach for interpreting the electronic wavefunctions of molecular clusters implying that cluster orbitals (CO) can be approximated by linear combinations of molecular orbitals (LCMO). Such CO LCMO scheme is no less founded than conventional MO LCAO one, since relative changes in the electron density distribution of individual molecules that accompany the formation of intermolecular bonds are definitely smaller than those typical of intramolecular chemical bonds. 

Results considered below are obtained in quantum chemical calculations carried out at the MP2 level with a 6-31G basis set augmented with diffuse and polarization functions on all nuclei (6-31++G(d,p)). This level of theory was repeatedly proved to be adequate for describing the systems of interest. All the configurations correspond to the energy minima, which is confirmed by the normal-coordinate analysis. Dissociation energies are estimated taking into account counterpoise corrections for the basis set superposition error, and the mean H-bond energy is found by dividing the total dissociation energy of a cluster (X$_n$ $\rightarrow n$X, where X = H$_2$O, HF, or NH$_3$) by the number of H-bonds. All calculations were carried out with the use of GAMESS~\cite{GAMESS} and Firefly~\cite{Firefly} program packages. 

\subsection{HF, H$_2$O, and NH$_3$ molecules: MO basis}

Orbitals of individual water, ammonia, and hydrogen fluoride molecules, which serve as the basis at the stage of constructing cluster orbitals, are well known and look as quite simple combinations of atomic orbitals. Let us denote orbitals of individual molecules as $\phi_i^f$, $\phi_i^w$, and $\phi_i^a$ in the case of hydrogen fluoride, water, and ammonia respectively, while those of clusters as $\varphi_i^f$, $\varphi_i^w$, and $\varphi_i^a$ \footnote{ Sometimes, superscripts are omitted when no confusion may arise (e.g., in figures).}. Being strongly localized $1s$ core orbitals of fluorine, oxygen, and nitrogen atoms respectively, $\phi_1^f$, $\phi_1^w$, and $\phi_1^a$  molecular orbitals are not of interest from the point of view of the formation of intermolecular bonds. Below, we briefly remind how the bonding orbitals of individual molecules look like omitting in the mathematical expressions those atomic orbitals, which contribute with negligible or much smaller weights, and dropping coefficients that depend on internuclear distances and particular functional form of the orbitals.

\begin{figure}[ht] \center
\includegraphics[width=0.8\textwidth]{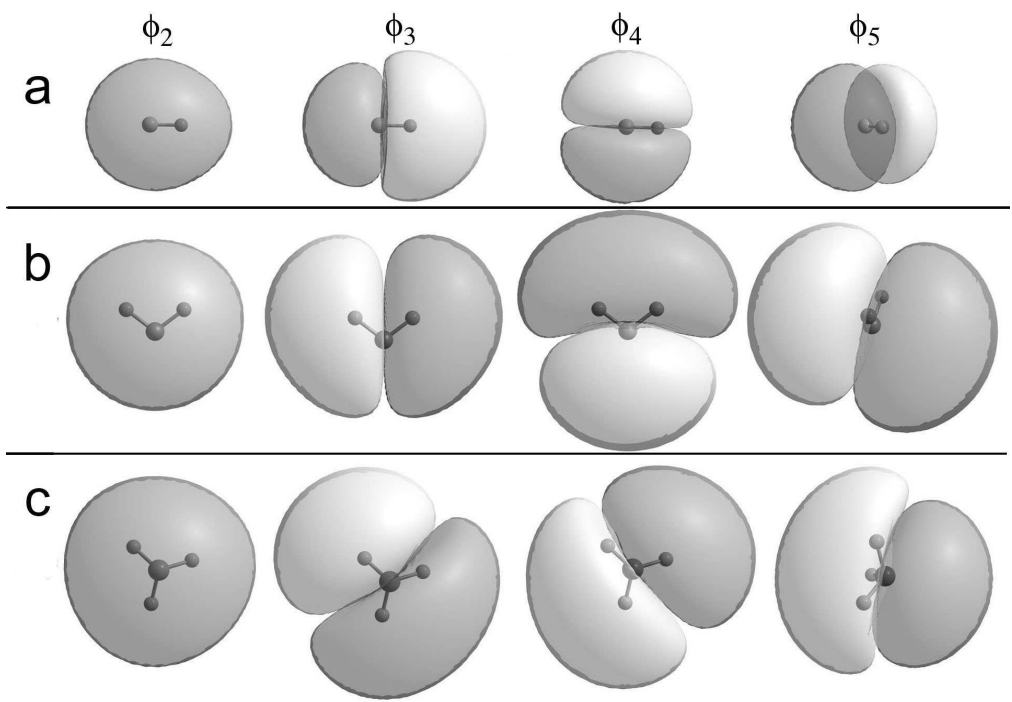}
\caption{Valence molecular orbitals of individual (a) HF, (b) H$_2$O, and (c) NH$_3$ molecules.} \label{fig1}
\end{figure}

In an HF molecule (Fig. \ref{fig1}a),\footnote{Figures \ref{fig1}, \ref{fig3}--\ref{fig5}, \ref{fig7}, and \ref{fig9}--\ref{fig11} show the surfaces of constant $\phi$ = 0.0015 or $\varphi$ = 0.0015 value, i.e., $\rho$ = 2.25$\cdot 10^{-6}$ au. The darker regions correspond to the positive values of $\phi$ or $\varphi$ function, while the lighter regions, to the negative values, so that their boundaries are the zero electron density surfaces. Orbitals in Figs. \ref{fig1} and \ref{fig3} are shown as transparent so that one can clearly see positions of nuclei.} chemical bonding is determined by two orbitals, one of which ($\phi_2^f$) is a linear combination of $2s$(F) and $1s$(H)\footnote{We use conventional notations for the atomic orbitals that involve shell numbers.} atomic functions, while the other ($\phi_3^f$) is a combination of $2p_z$(F) and $1s$(H) functions ($Oz$ axis is oriented along the F--H line). The residual four valence electrons are described by two degenerate orbitals, namely, $\phi_4^f = 2p_x$(F) and $\phi_5^f = 2p_y$(F), and represent the electron density that can be referred to as {\textquotedblleft}free{\textquotedblright}, i.e., uninvolved in interatomic binding. 

In an H$_2$O molecule (Fig. \ref{fig1}b), the lowest valence orbital ($\phi_2^w$) is a linear combination of $2s$(O), $1s$(H$^1$), and $1s$(H$^2$) atomic functions (where superscript enumerates protons). Two next orbitals are bonding in O--H$^1$, O--H$^2$ and H$^1$\dots H$^2$ domains respectively: $\phi_3^w = 2p_x$(O) + ($1s$(H$^1$) -- $1s$(H$^2$)) ($xOz$ is the nuclear plane of the molecule, and $Oz$ axis is the symmetry axis of the molecule) and $\phi_4^w = 2p_z$(O) + ($1s$(H$^1$) + $1s$(H$^2$)). Finally, a lone pair, which represents the free density of the molecule, is described by $\phi_5^w = 2p_y$(O) orbital. 

In an NH$_3$ molecule (Fig. \ref{fig1}c), the lowest bonding orbital is $\phi_2^a = 2s$(N) + ($1s$(H$^1$) + $1s$(H$^2$) + $1s$(H$^3$)). Two next degenerate orbitals ($\phi_3^a$ and $\phi_4^a$) provide the electron binding in the N--H$^1$, N--H$^2$, and N--H$^3$ domains: $\phi_3^a = 2p_x$(N) + ($1s$(H$^2$) + $1s$(H$^3$) -- $1s$(H$^1$)) and $\phi_4^a = 2p_y$(N) + ($1s$(H$^2$) -- $1s$(H$^3$)) ($Oz$ axis coincides with the triad axis of the structure, and $Ox$ axis coincides with the projection of N--H$^1$ bond on the orthogonal plane). The residual valence orbital is also bonding: $\phi_5^a = 2p_z$(N) -- ($1s$(H$^1$) + $1s$(H$^2$) + $1s$(H$^3$)). Thus, by contrast to water and hydrogen fluoride molecules, there is no electron uninvolved in intramolecular bonds in ammonia; only about a half of the electron density described by the latter orbital and localized outside the NH$_3$ pyramid can conditionally be referred to as free. 

Thus, if one supposes that the main contribution to the formation of hydrogen bonds is provided by the redistribution of the {\textquotedblleft}free{\textquotedblright} electron density of the molecules, the most spatially flexible systems of H-bonds with the highest bond energies should be observed in the case of HF molecules and the smallest flexibility and the weakest bonds, in the case of NH$_3$. Let us check the supposition by considering some typical small (HF)$_n$, (H$_2$O)$_n$, and (NH$_3$)$_n$ clusters.

However, before proceeding to particular examples, it is worth noting that intramolecular bonds form due to the overlap of the frontier orbitals of interacting particles (atoms or functional groups). These are both highest occupied and lowest unoccupied molecular orbitals or singly occupied molecular orbitals of both reagents; and in the resulting system (in the ground electronic state), bonding orbitals are usually occupied, while antibonding ones are unoccupied. When one deals with clusters composed of closed-shell molecules, the situation is different: the redistribution of the electron density is due to the overlap of the occupied orbitals of the particles, so that the resulting functions, which describe the electron density distribution in the cluster, are of both kinds, namely bonding and antibonding. Hence, the total binding effect is much smaller compared to typical interatomic bonds and depends on the balance between bonding and antibonding contributions in each intermolecular region.

\subsection{(HF)$_n$ clusters}

\begin{figure}[ht] \center
\includegraphics[width=0.55\textwidth]{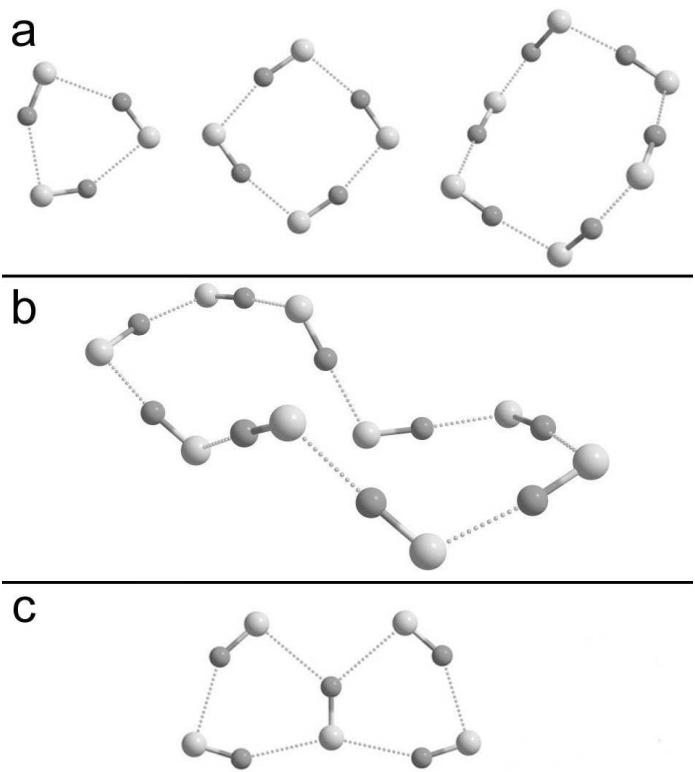}
\caption{Typical structures of (HF)$_n$ clusters: (a) plain rings; (b) extended folded rings; and (c) openwork species composed of trimolecular rings.} \label{fig2}
\end{figure}

HF molecule typically forms two bonds with the neighbors ($da$ coordination), though there are situations when either hydrogen or fluorine atom or both of them are involved in two bonds with the neighboring molecules. Stable structural units are molecular rings (Fig. \ref{fig2}a) where molecules are arranged in such a way that H\dots F--H angles are close to 90$^{\circ}$, while dihedral F--H\dots F--H angles are arbitrary. When no less than six molecules are joined in a ring, it looks like a folded sheet or strip composed of rectangular segments (Fig. \ref{fig2}b). Openwork species, in which three-membered molecular rings have common HF edges, are also possible (Fig. \ref{fig2}c). Larger rings that involve three-coordinated molecules can be joined in nonplanar openwork species. Such building blocks enable one to construct a two-dimensional network, but any attempt to build a continuous three-dimensional bonded system fails. Whenever relatively small ring-like fragments can separate, they do. Let us try to understand the peculiarity by considering bond energies typical of different coordination neighborhoods. 

When the number of molecules in (HF)$_n$ rings (Fig. \ref{fig2}a) increases from 3 to 5, the mean H-bond energy changes from 5.2 to 8.6 kcal/mol. In extended folded rings similar to that shown in Fig. \ref{fig2}b, a mean bond energy is about 8.4 kcal/mol. Two parallel molecular rings have no noticeable attraction to each other. Species based on trimolecular rings with common HF edges (Fig. \ref{fig2}c) are less stable than plain rings: a mean H-bond energy is about 4.2 kcal/mol. With an increase in the number of molecules, which form more than two hydrogen bonds, the mean bond energy decreases to about 4 kcal/mol. A mean bond energy per molecule is higher, about 5 kcal/mol, but still substantially lower than that in plain molecular rings. Bonds in chains are characterized by intermediate energies, e.g., about 5.9 and 6.4 kcal/mol in four- and five-membered chains. 

Thus, knowing the bond energies in diverse structural units, one can predict most probable kinds of H-bonded systems. However, the information cannot clarify the underlying reasons of stability. For example, the difference between the bond energies in molecular rings and chains (2--2.5 kcal/mol) seems too large for the systems composed of molecules that have no more than two neighbors. At the same time, when the number of H-bonds formed by a molecule increases to three or four, their energies decrease despite nearly the same change in the atomic charges, which a re conventionally thought of as correlating with the bond strengths. hence, the bond energies depend on the character of the electron density distribution rather than its integral value in the corresponding spatial regions. Peculiarities of the distribution can be judged from the character of cluster orbitals.

\begin{figure}[ht] \center
\includegraphics[width=0.7\textwidth]{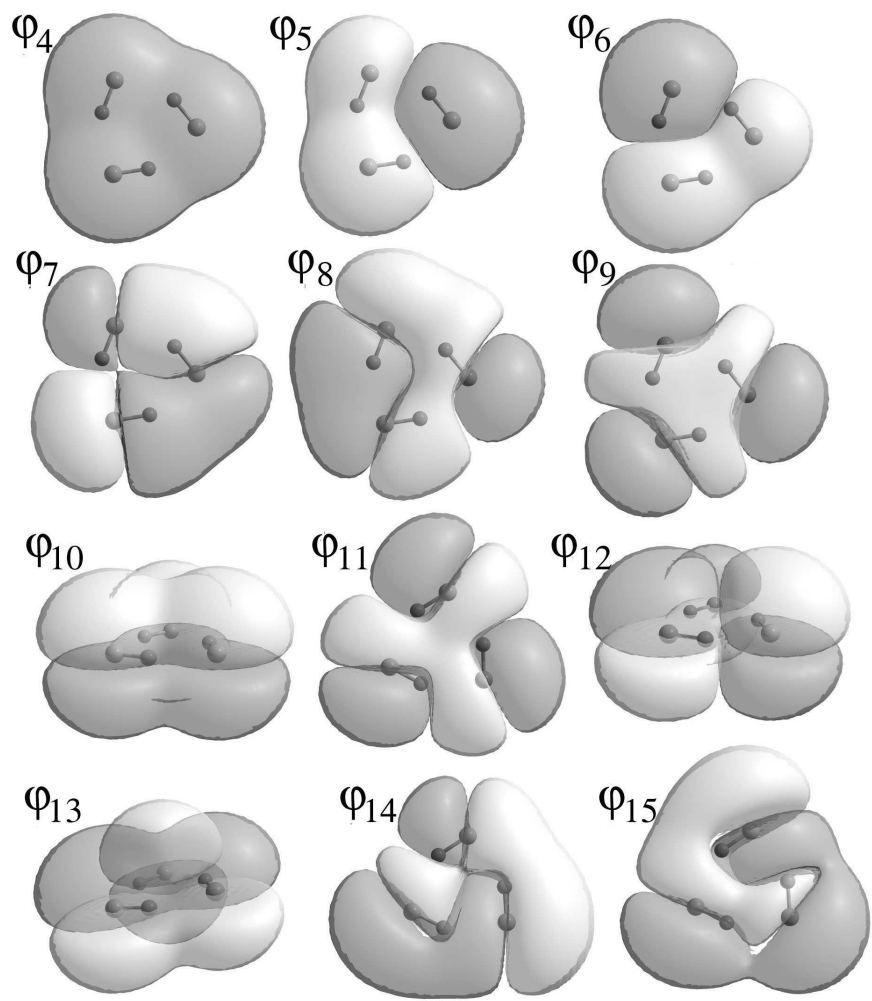}
\caption{Valence orbitals of (HF)$_3$ cluster.} \label{fig3}
\end{figure}

Let us start with (HF)$_3$, which is the smallest ring-like structure. Even a prompt glance distinguishes two kinds of cluster orbitals (Fig. \ref{fig3}). The lowest $\varphi_m^f$ orbitals with $m$ = 4, 5, and 6 (three core functions are again omitted) are obviously linear combinations of $\phi_2^f$  molecular orbitals and have definitely a $\sigma$-kind. To the approximations of all the residual cluster orbitals, $p$-functions of fluorine atoms noticeably contribute. Six of them, namely, $\varphi_m^f$ with $m$ = 7, 8, 9, 11, 14, and 15, are also of $\sigma$-kind and describe either electronic binding or antibinding in the H-bond domains. Let us denote the plane of fluorine nuclei as $xOz$; then, these cluster orbitals can be considered as linear combinations of $\phi_3^f$ and $\phi_4^f$ molecular orbitals. Note that the residual cluster orbitals with $m$ = 10, 12, and 13 have definite $\pi$-character and are approximated by linear combinations of $\phi_5^f$  molecular orbitals. These cluster orbitals have much in common with typical $\pi$-system functions of a conjugated hydrocarbon. The sole peculiarity is as follows. In hydrocarbon rings, maxima of the $\pi$ functions are at carbon nuclei, while minima and zeros are at the midpoints of C--C bonds. In (HF)$_3$ cluster, the large-value parts of orbitals correspond to H--F regions with maxima at F nuclei and minima in H\dots F domains close to bridge proton positions. Thus, the orbitals describe a \textit{delocalized $\pi$-binding} of fluorine atoms; and H atoms here act as auxiliary bridge sites.

All $\sigma$ and $\pi$ cluster orbitals remain bonding in the regions of intramolecular H--F bonds, but can be either bonding or antibonding in the hydrogen-bond domains. For example, of the three cluster orbitals that can be constructed (taking into account the general $\mathbb{C}_3$ symmetry of the cluster) from three $\phi_2^{f(i)}$ molecular orbitals ($i$ superscript enumerates molecules), only one ($\varphi_4^f$) is bonding, while two residual have prevailing antibonding character: 
\begin{equation*} 
\begin{split}
\varphi_4^f &= \phi_2^{f(1)} + \phi_2^{f(2)} + \phi_2^{f(3)} \\
\varphi_5^f &= 2\phi_2^{f(3)} - \phi_2^{f(1)} - \phi_2^{f(2)} \\
\varphi_6^f &= 2\phi_2^{f(1)} - \phi_2^{f(2)} - \phi_2^{f(3)}. \\
\end{split}
\end{equation*}

Insofar as $\sigma$ cluster orbitals are linear combinations of either $\phi_2^f = 2s$(F) + $1s$(H) or $\phi_3^f = 2p_z$(F) + $1s$(H) bonding orbitals and $\phi_4^f = 2p_x$(F) lone pair orbitals, whose axes are mutually orthogonal, HF molecules tend to be arranged in clusters in such a way that H\dots F--H angles are close to 90$^{\circ}$. Planar configuration of (HF)$_3$ ring also seems natural, since the additional $\pi$ binding is provided by the overlap of  $\phi_5^f = 2p_y$(F) molecular orbitals whose axes are normal to the plane of $\sigma$-binding. 

\begin{figure}[ht] \center
\includegraphics[width=0.7\textwidth]{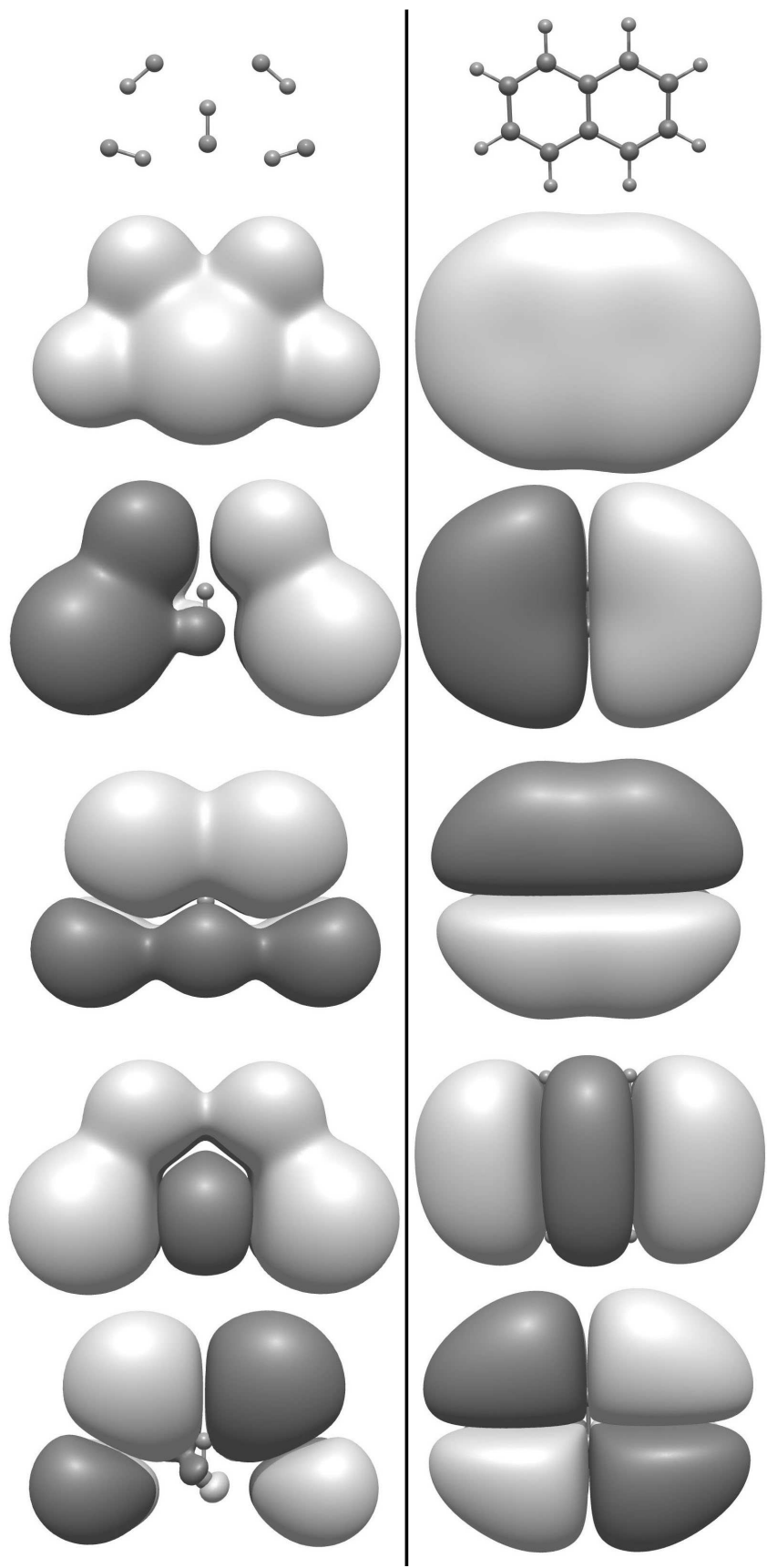}
\caption{$\pi$-orbitals of (a) (HF)$_5$ openwork structure and (b) naphthalene.} \label{fig4}
\end{figure}

The same character is typical of the four-membered (HF)$_4$ ring and openwork (HF)$_n$ clusters, in which trimolecular rings are joined via common HF edges (see Fig. \ref{fig2}). On the whole, changes in the orbital picture that take place with an increase in the ring size or upon fusion of the rings are similar to those observed in the case of hydrocarbon rings when the number of C nuclei in the ring skeleton increases or when rings of the same size are joined via edges. An interesting illustration is given by Fig. \ref{fig4} where five occupied $\pi$ orbitals of (HF)$_5$ openwork structure are compared to five $\pi$ orbitals of naphthalene. In the case of fused rings, the electron density, which is {\textquotedblleft}contributed{\textquotedblright} by a common edge HF molecule to the formation of hydrogen bonds, is nearly twice as small as the density, which provides binding of molecules in plain rings. As a result, the energies of such bonds are noticeably lower. 

\afterpage{\clearpage}

The similarity of the orbital pictures of small (HF)$_n$ rings and aromatic hydrocarbons does not apply to chain-like structures. In planar molecular (HF)$_4$ or (HF)$_5$ chains, there are both $\sigma$ and $\pi$ cluster orbitals, but each $\pi$ orbital describes chiefly the electron binding of some two neighboring molecules. Thus, by contrast to rings, $\pi$ orbitals in the chains are noticeably localized. This is clearly reflected in the internuclear distances. For example, in (HF)$_4$ chain, the increase in H--F distances in the central molecules (compared to the individual molecule) is about 0.02 \AA\, while that in the terminal molecules is no more than 0.01 \AA. Internuclear distances are not equal and fall in a range from 1.636 to 1.714 \AA. At the same time, in the ring of the same molecular size, the increase in the intramolecular H--F distances is 0.035 \AA, and the intermolecular H\dots F distances are all the same, about 1.534~\AA.  

The more localized $\pi$ bonding orbitals, the weaker the bonds; and the difference in energy is substantial. When a zigzag tetra- or pentamolecular chain closes in a ring, the absolute energy decrease is 17--20 kcal/mol, which is nearly thrice as large as an individual H-bond energy in the ring. This means that such closing is accompanied by the additional stabilization of the already existing hydrogen bonds, which is reached due to the appearance of a delocalized conjugated $\pi$ system.

To sum up, the combination of $\sigma$ and $\pi$ binding explains the relative stability of small planar (HF)$_n$ clusters. Planarity is the characteristic feature of conjugated hydrocarbons. However, as was mentioned above, when there is no less than six molecules in an (HF)$_n$ ring, it is no longer planar (even at $n$=5, there is already a small deviation from planarity). It is folded and consists of rectangular segments. This peculiarity agrees with the trend of HF molecules to form sequences of mutually orthogonal hydrogen bonds, which is impossible in the case of planar structures. The orthogonality is determined by $\sigma$ binding. Hence, we can state that this kind of electronic binding is present in all the clusters and determines the spatial arrangement of molecules. Then, what is the role of $\pi$ orbitals and do they exist in nonplanar systems? It turns out that they do, but in some new variant, which is unusual for any person who has some acquaintance with organic chemistry.

\begin{figure}[ht] \center
\includegraphics[width=0.7\textwidth]{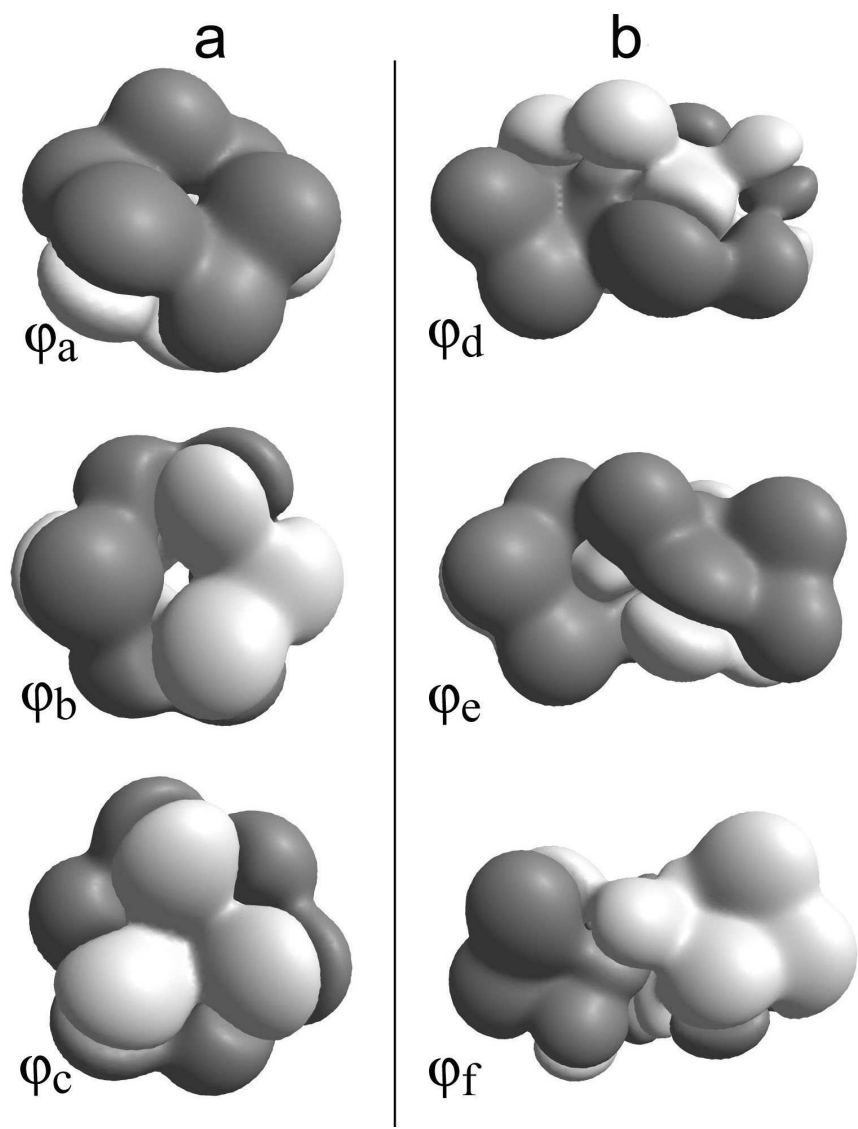}
\caption{Typical valence $\pi$-orbitals of nonplanar (HF)$_n$ clusters: (a) $n$ = 6 and (b) $n$ = 9.} \label{fig5}
\end{figure}

As an example, let us consider $\pi$ orbitals of a ring-like (HF)$_6$ cluster (Fig. \ref{fig5}a). The lowest bonding orbital of this kind ($\varphi_a$) has much in common with usual $\pi$ orbitals of conjugated hydrocarbons with a sole difference: with an increase in the number of molecules, minima of the function (more accurately, (3, --1) points) become closer to the position of the bridge proton. (A similar trend is observed in the case of $\sigma$ cluster orbitals.) All the residual $\pi$ cluster orbitals dramatically differ from those of conjugated hydrocarbons and planar (HF)$_n$ clusters. Due to the planar configuration of all the latter systems, there were definite charge-alternation surfaces in all of them located in the H-bond domains. 

In (HF)$_6$ cluster, constant-sign parts of the orbitals look as closed rings, and the rings are connected as links of a chain ($\varphi_b$ in Fig. \ref{fig5}a). The highest orbital resembles a chain rivet ($\varphi_c$  in Fig. \ref{fig5}a), as if in a six-link ring of one sign, a rod of the other sign is inserted, and both its ends are riveted to triangles. Thus, the antibonding regions are located either in the vicinity of bridge protons or in the central part of the cluster. And electrons of all six $\pi$ orbitals, each of which is delocalized over the whole hexamolecular ring, contribute to the intermolecular binding.

When the number of molecules is increased above eight, and pronounced folded structures appear, the orbital picture becomes more complex (Fig. \ref{fig5}b). Even the lowest $\pi$ orbital looks already as two interlaced rings (of different signs) and resembles a typical $\pi$ bonding orbital only in a two thirds of the ring ($\varphi_d$). In the next orbitals, which are more and more complex interlacements of multifold bent rings, the sign-alternation regions already coincide with the H-bond domains, so that the electronic binding in certain H\dots F regions is weaker and, at the same time, there is no general pronounced binding of all the molecules. 

In a general case, if we look at a folded structure from the side, we can see a zigzag; and the more strongly internally bound subsystems correspond to separate periods of the zigzag. For example, in (HF)$_9$, tetra- and pentamolecular subsystems ($\varphi_e$ and $\varphi_f$ in Fig. \ref{fig5}b) can be distinguished. Quantitatively, this separation of quasi-aromatic segments in extended folded structures is reflected by the mean H-bond energy of the cluster, which decreases from 9.2 kcal/mol in (HF)$_6$ to 8.3 kcal/mol in (HF)$_9$. 

The discovered peculiarities of electronic binding between HF molecules show that hydrogen fluoride should form structures composed of lamellar domains, the most stable building units of which should be planar or folded (depending on the molecular size) rings. In the rings, molecules are joined via electronic bonds, in which $\sigma$ and $\pi$ contributions are clearly distinguishable, and $\pi$ binding system is delocalized conjugated. With an increase in the molecular size of the ring, which is accompanied by the formation of nonplanar structures, the H-bond energies increase (to a certain limit) due to the stronger $\pi$ binding. When the ring becomes too large, and no effective conjugated system of bonds can be formed, strongly bound subsystems are separated. 

Note that two parallel zigzag chains located at a distance of about 2.5 \AA\ (equal to F--H\dots F distance) should form a similar $\pi$-bonded agglomerate. However, in the absence of terminal closing molecules, which keep the chains at a constant distance, such agglomerate should exist only at a low temperature. With an increase in temperature, it should probably transform in several rings arranged successively. At the same time, antiparallel zigzag chains can easily be interconnected via properly reoriented molecules, so that either one extended ring or a similar succession of rings can appear. 

These peculiarities explain why hydrogen fluoride is a mobile and volatile liquid.

Thus, in any case, we deal with a coherent binding of all the molecules or some their large subgroups rather than simply binding of molecular pairs. Let us now see whether a similar conjugation exists in clusters composed of water and ammonia molecules, the ensembles of which cannot have planar configurations, but X--H\dots X--H\dots bond sequencies (X = O or N) can form planar rings in some cases. Note that both H$_2$O and NH$_3$ molecules can form up to four {\textquotedblleft}normal{\textquotedblright} hydrogen bonds with the neighbors, which makes them more promising as building blocks of three-dimensional networks. Let us start with ammonia clusters, which should be in some sense opposite to hydrogen fluoride ones, since the {\textquotedblleft}free{\textquotedblright} electron density of ammonia molecules is the lowest (compared to HF and H$_2$O).

\subsection{(NH$_3$)$_n$ clusters}

\begin{figure}[ht] \center
\includegraphics[width=0.9\textwidth]{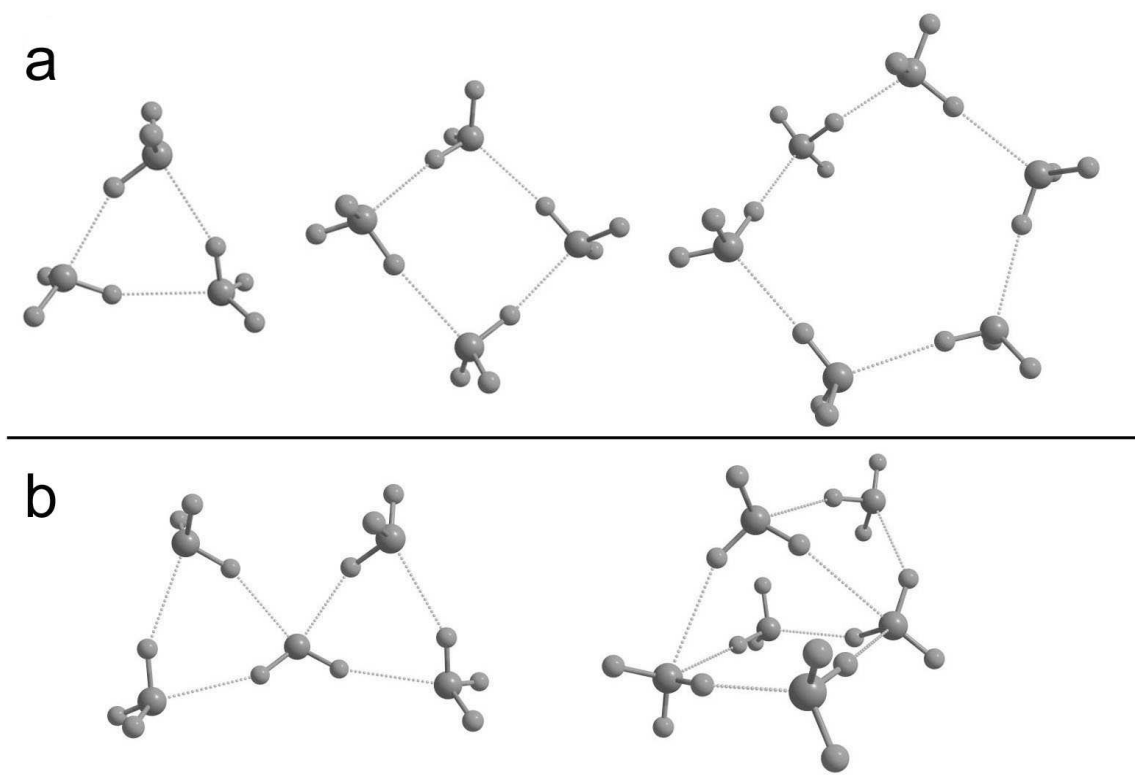}
\caption{Typical structures of (NH$_3$)$_n$ clusters: (a) plain rings and (b) joined or fused tri- and tetramolecular rings.} \label{fig6}
\end{figure}

Ammonia molecule can normally form three hydrogen bonds as a proton donor and one as an acceptor. When its coordination neighborhood is of $da$ kind (Fig. \ref{fig6}a), i.e., molecules form a ring-like structure, the mean bond energy is nearly by half smaller compared to that in (HF)$_n$ clusters. For example, when the number of molecules in a ring increases from three to six, the energy increases from 3.6 to 4.5 kcal/mol. When small molecular rings are joined in branched or cage-like structures via molecules whose nitrogen atoms are involved in two hydrogen bonds with the neighbors (Fig. \ref{fig6}b), the mean bond energy is lower: about 3.1 kcal/mol. In such systems, similarly to (HF)$_n$ clusters, the bond energy per molecule is higher (about 3.8--4.0 kcal/mol), but still lower than that in plain rings. 

The formation of hydrogen bonds between ammonia molecules is accompanied by such electron density redistribution that practically does not touch those hydrogen atoms, which are not involved in the bonds formed. The extent of integral charge redistribution is comparable to that in (HF)$_n$ clusters, but the bond energies in the systems differ nearly by half. This indicates again that integral electron density characteristics can by no means be unique indicators of the H-bond strength. 

Orbital picture of (NH$_3$)$_3$ is similar to that obtained in the case of hydrogen fluoride trimer. Twelve valence cluster orbitals ($\varphi_m^a$) can be divided in four threes ($m$ = 4, 5, 6; 7, 11, 12; 8, 9, 10; 13, 14, 15), each of which is represented by linear combinations of the molecular orbitals of the same kind, namely, $\phi_2^a$, $\phi_3^a$, $\phi_4^a$, and $\phi_5^a$ respectively. Here, $\phi_k^a$  molecular orbitals with $k$ = 2, 3, or 5 provide combinations of $\sigma$ kind, whereas overlapped $\phi_4^a$ orbitals give $\pi$ functions, one of which ($\varphi_8^a$) is a typical bonding orbital, while two residual ($\varphi_9^a$ and $\varphi_{10}^a$) are degenerate and describe electron bonding of some two molecules and antibonding in two residual pairs.

Thus, $\sigma$-binding of ammonia molecules is provided by the "free" electron density of the molecules ($\phi_5^a$ functions) and the electrons described by one of two degenerate molecular orbitals ($\phi_3^a$). Electrons that were described initially by $\phi_4^a$ orbitals (orthogonal to $\phi_3^a$ functions) represent the conjugated $\pi$ system of the molecular ring. Orientation of $\phi_5^a$ orbital along the triad axis of NH$_3$ pyramid determines the preferable linear configuration of N\dots H--N fragments, i.e., the linearity of hydrogen bonds; whereas the original degeneracy of $\phi_3^a$ and $\phi_4^a$ orbitals of individual molecules leads to the arbitrary orientation of N$^{(1)}$H$_3$ pyramid with respect to N$^{(1)}$--H line in (H$_2$)N$^{(1)}$--H\dots N$^{(2)}$(H$_3$) hydrogen-bonded pair and an arbitrary rotation angle of N$^{(2)}$H$_3$ pyramid about its triad axis. Accordingly, possible mutual orientations of NH$_3$ pyramids in (NH$_3$)$_n$ clusters can be very different. The relative closeness of HNH valence angle in the individual ammonia molecule (ca. 113$^{\circ}$) to the planar angle in an equilateral triangle provides the formation of such (NH$_3$)$_3$ structure where (N--H)$_3$ bond ring is planar and have much in common with the (HF)$_3$ ring. The difference is the substantially lower electron density in the H-bond domains, which is determined by the fact that it is electrons, which provide the existence of intramolecular N--H bonds, rather than lone pair electrons that contribute both to $\sigma$ and $\pi$ binding. As a result, bond energies are much smaller: ~3.5 kcal/mol in (NH$_3$)$_3$ versus ~5.3 kcal/mol in (HF)$_3$. 

In a tetramolecular ring, the orbital picture is principally similar; though, due to a small deviation of (N--H)$_4$ ring from planarity, nonbonding and antibonding $\pi$ orbitals are already slightly similar to interlaced rings of different sign. However, the electron density distribution within such rings is substantially nonuniform, with constrictions in the H-bond domains close to proton positions. 

\begin{figure}[ht] \center
\includegraphics[width=0.75\textwidth]{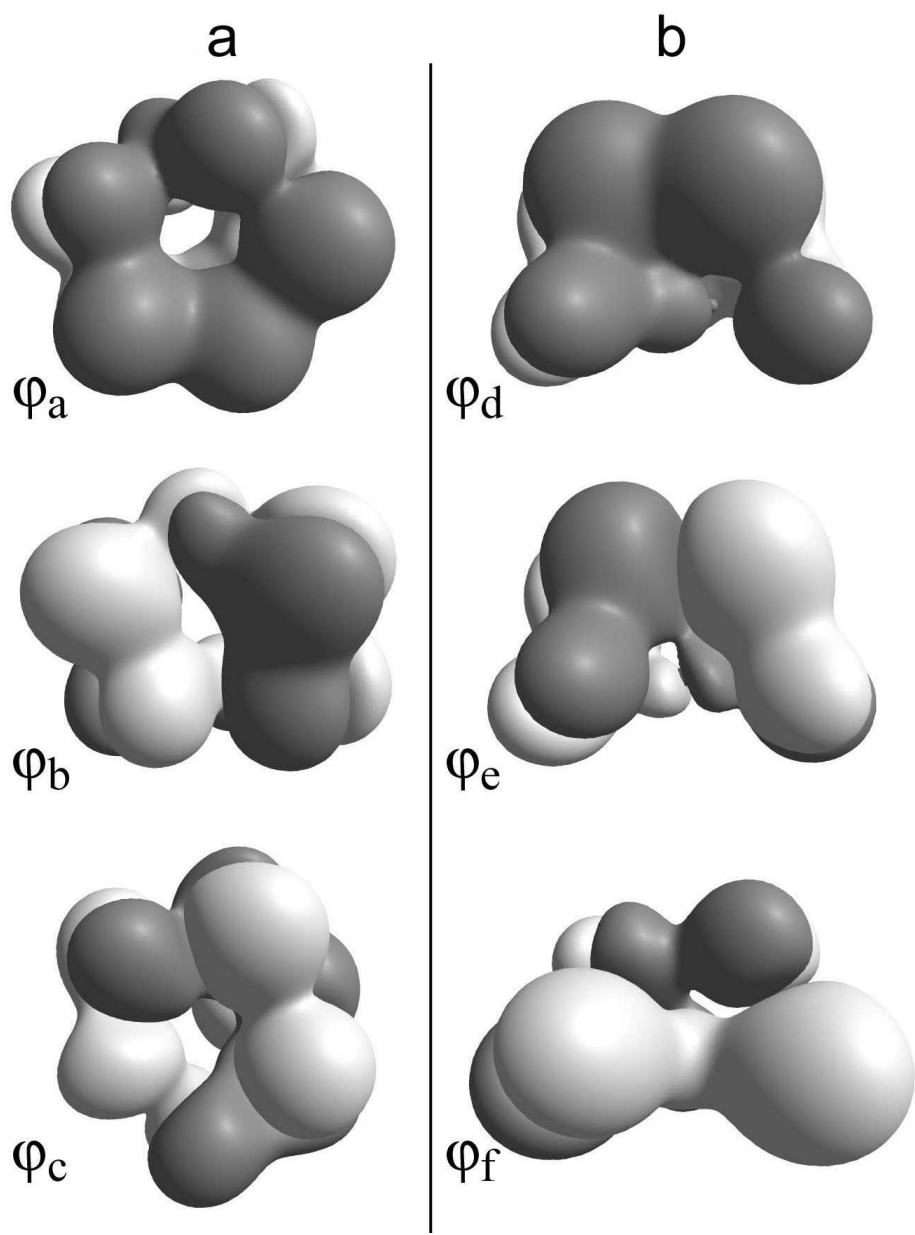}
\caption{Typical valence $\pi$ orbitals of (NH$_3$)$_n$ clusters: (a) $n$ = 6 and (b) $n$ = 5.} \label{fig7}
\end{figure}

In the case of a hexamolecular (NH$_3$)$_6$ ring, which has a bath-like configuration, $\pi$-orbitals look like constant-sign rings interlaced in a complex way (Fig. \ref{fig7}a); and the rings are broken except for the sole orbital, $\varphi_a$. The latter statement is conditional, since in a constant-density map with a boundary value smaller than $10^{-6}$ au, the rings will be seen as closed, but such low electron density provides actually no binding in the corresponding spatial regions. As can be seen, delocalized $\pi$ bonds exist within open three- or five-membered rings. 

A similar picture is observed in (NH$_3$)$_5$ cluster when it has a configuration of two trimolecular rings joined via a common molecule (Fig. \ref{fig7}b). The latter molecule has a $ddaa$ coordination, so that despite its low {\textquotedblleft}free{\textquotedblright} density it acts as an acceptor of two protons. Even the lowest $\pi$ orbital of the cluster ($\varphi_d$ in Fig. \ref{fig7}b) describes a noticeable electron binding of only four molecules, i.e., except for the tetracoordinated central one. The residual $\pi$ cluster orbitals determine binding within either dimeric fragments or a dimer and a chain-like trimer ($\varphi_e$ and $\varphi_f$ in Fig. \ref{fig7}b). In all the cases, the electron density within the spatial region of the central molecule and those its hydrogen bonds, in which it acts as an acceptor of protons (i.e., an electron density donor), is the lowest.

Thus, the very low electron density, which can be contributed by an ammonia molecule to the formation of $\pi$ system can provide additional stabilization of only small rings (composed of three to five molecules) or small chain-like fragments within larger clusters. In the regions of tetracoordinated molecules, there is no continuous delocalized $\pi$ system that could stabilize the joint. Therefore, it is highly probable that the structure reorganization in such regions can be initiated by a very small external perturbation (or energy fluctuation). The result will be the breakage of one or two H-bonds formed by the central molecule, and the appearance of bent chain-like fragments only partially stabilized by $\pi$ binding. 

Ring-like species with quasi-aromatic $\pi$ systems, which were found to be most stable in the case of (HF)$_n$ clusters, are very small (three- to four-molecular) and statistically rare objects in ammonia molecular ensembles. This peculiarity explains, particularly, the much lower boiling point of liquid ammonia compared to hydrogen fluoride. 

Now, let us turn to water clusters whose molecules have free electron density, which is noticeably larger than that of ammonia molecules, but smaller than that of hydrogen fluoride.

\subsection{(H$_2$O)$_n$ clusters}

\begin{figure}[ht] \center
\includegraphics[width=0.8\textwidth]{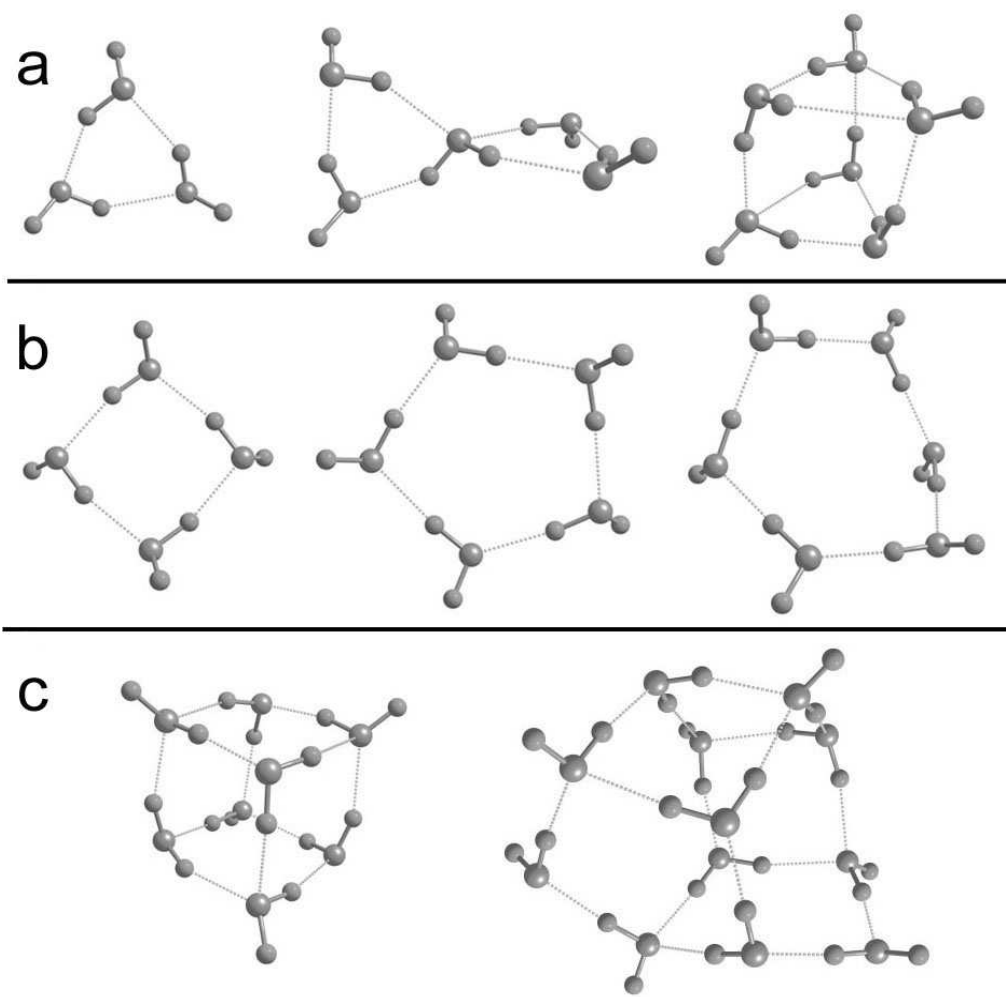}
\caption{Typical structures of (H$_2$O)$_n$ clusters: (a) composed of or involving trimolecular rings; (b) plain rings with nearly straight hydrogen bonds; and (c) cages composed of fused tetra-, penta-, and hexamolecular rings.} \label{fig8}
\end{figure}

The mean bond energy in (H$_2$O)$_n$ clusters is known to increase with an increase in $n$ up to a certain limit, which is reached at a number of molecules about twenty. Here, people typically speak about the bond energies per molecule. However, the interaction energy per H-bond taking into account the neighborhood of molecules is more informative. Let us distinguish the following coordination kinds: $da$, $dda$, $daa$, and $ddaa$. If all molecules are involved in trimolecular rings, then, practically irrespectively of their particular local neighborhoods ($da$ in a ring-like trimer or $da$ and $ddaa$ in a pentamer where two trimolecular rings are joined via a common molecule, Fig. \ref{fig8}a), the mean H-bond energy is 5.0 kcal/mol. In (H$_2$O)$_n$ rings of the larger size (Fig. \ref{fig8}b) where H-bond angles are closer to 180$^{\circ}$ and all molecules have $da$ coordination, the interaction energy per bond is 6.5, 6.8, and 6.9~kcal/mol at $n$ = 4, 5, and 6 respectively. In (H$_2$O)$_6$ prism, (H$_2$O)$_8$ cube, and (H$_2$O)$_{12}$ cage, which are composed of fused three-, four-, five-, and six-membered molecular rings where tricoordinated molecules prevail (Fig. \ref{fig8}a, c), a mean H-bond energy depends on the steric factor, being 4.7, 5.7, and 6.0 kcal/mol, respectively. Thus, with an increase in the number of hydrogen bonds formed by the same molecule, the mean bond energy slightly decreases, while with an increase in the size of the ring, it increases, so that the H-bond energy per molecule increases with an increase in the molecular size of the cluster, being 6.9, 8.5, and 8.6 kcal/mol in the aforementioned (H$_2$O)$_n$ clusters with $n$ = 6, 8, and 12.

Thus, water molecules seem to be most promising for the formation of three-dimensional bond networks. Energies of individual bonds in plain ring-like (H$_2$O)$_n$ clusters are higher than those in (NH$_3$)$_n$ clusters and lower than those in (HF)$_n$. However, it is water clusters solely where the mean bond energy remains sufficiently high when the coordination number of the molecule increases from two to four; and the H-bond energy per molecule even increases (up to a certain limit). 

By contrast to both ammonia and hydrogen fluoride, water molecule has no degenerate orbitals. Recall that there are two degenerate orbitals in HF that describe the electron density of lone pairs; and there are two degenerate orbitals in NH$_3$ molecule that describe N--H binding. When molecules of both latter kinds form clusters, electrons of one of two degenerate orbitals contribute to $\sigma$ binding, while those of the other, to the formation of $\pi$-system. Orbitals of H$_2$O molecule are all nonequivalent in energy and electron density distribution. Therefore, the orbital picture should strongly depend on the arrangement of molecules, namely, on the relative orientations of their nuclear planes.

Nevertheless, based on the orbital picture of individual water molecule and the discovered peculiarities of the intermolecular bonds in (HF)$_n$ and (NH$_3$)$_n$ clusters, we can expect that $\sigma$ binding in (H$_2$O)$_n$ clusters should chiefly be provided by the overlap of intramolecular bonding orbitals, namely, $\phi_3^w = 2p_x$(O) + ($1s$(H$^1$) -- $1s$(H$^2$)) and  $\phi_4^w = 2p_z$(O) + ($1s$(H$^1$) + $1s$(H$^2$)), while lone pair electrons should provide $\pi$ binding. It is quite possible also that lone pair electrons can be involved in the formation of $\sigma$ bonds, while the overlap of intramolecular bonding orbitals can additionally contribute to the formation of $\pi$ system. In the molecule that acts as a proton donor, electrons described by $\phi_3^w$ and $\phi_4^w$ determine $\sigma$ binding. In the proton-acceptor molecule, diverse MOs can contribute to $\sigma$ binding depending on the mutual orientation of the nuclear planes of the molecules. Such ambiguity makes it more difficult to predict optimal mutual orientations of water molecules in clusters and the preferable coordination of molecules; but, at the same time, it determines the larger flexibility of the electron density distribution between $\sigma$ and $\pi$ systems.

\begin{figure}[ht] \center
\includegraphics[width=0.9\textwidth]{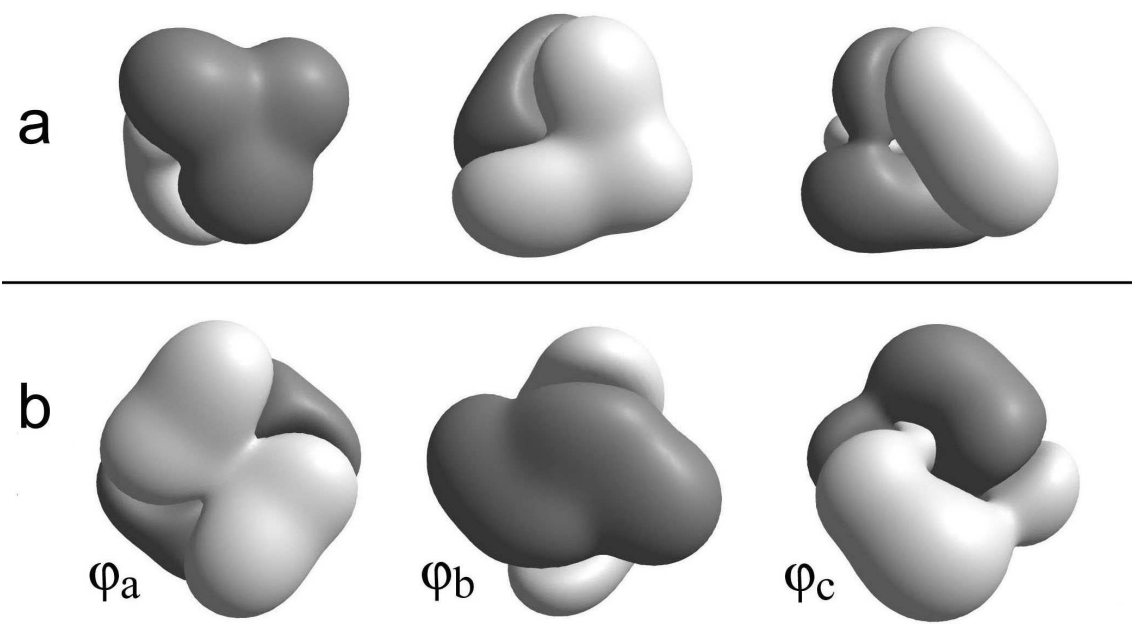}
\caption{Some valence $\pi$ orbitals of ring-like (H$_2$O)$_n$ clusters: (a) $n$ = 3 and (b) $n$ = 4.} \label{fig9}
\end{figure}

Let us start (as before) with the smallest ring-like cluster, namely trimer. Recall that both in (HF)$_3$ and (NH$_3$)$_3$, we could definitely distinguish $\sigma$ and $\pi$ cluster orbitals, and the latter orbitals were very similar to bonding and antibonding orbitals of a conjugated hydrocarbon. What do we see in the case of water trimer? Figure \ref{fig9}a shows three most interesting valence orbitals of the cluster. Their approximations involve $\phi_4^w$ and $\phi_5^w$ molecular orbitals and they look like twisted $\pi$ orbitals. They are twisted in such a way that within each O\dots H domain they have definitely $\pi$ kind. Alternation-sign surfaces of the orbitals always pass in close vicinity to bridge protons. 

The orbital picture of the ring-like water tetramer also represents an intermediate situation between those typical of planar and non-planar systems (Fig. \ref{fig9}b). There is an orbital, which resembles a nonbonding $\pi$ orbital of a planar four-membered ring ($\varphi_a$). Another one ($\varphi_b$) is a $\pi$-bonding orbital. Two residual orbitals are similar to those observed in non-planar clusters of hydrogen fluoride and ammonia, namely, look like interlaced constant-sign rings (Fig. \ref{fig9}b shows one of them, $\varphi_c$), where the sign-alternation surfaces pass in close vicinity to bridge-proton positions. It is worth noting that, despite the visible resemblance, the orbitals have a definite peculiarity. Zero-density surfaces are very complex, and in many cases, they divide two bonding domains, namely, intramolecular O--H and the neighboring intermolecular H\dots O domains, so that the generally $\pi$-kind orbital that spans the whole cluster has a mixed $\pi$/$\sigma$ character in the H-bond domains.

In ring-like pentamer and hexamer, the situation is principally similar. There are five and six cluster orbitals of generally $\pi$ kind respectively. Character of the electron density distribution described by any of the orbitals can be named intermediate between those typical of hydrogen fluoride and ammonia clusters. This means that there are bonding and antibonding domains, but the constant-sign rings always look like bended or twisted doughnuts with tight constrictions. In fact, this observation strongly correlates with the {\textquotedblleft}amount{\textquotedblright} of free electron density that can be localized in the H-bond domains. 

This result explains also why the mean bond energies in ring-like water clusters are intermediate between those in hydrogen fluoride and ammonia clusters of the same size and why the difference in bond energies between hydrogen fluoride and water clusters increases with an increase in the number of molecules involved in the ring. It is about 0.1, 1.0, 1.6, and 2.4 kcal/mol in the case of trimers, tetramers, pentamers, and hexamers, respectively. Thus, the most stable are rings composed of HF molecules. 

\begin{figure}[ht] \center
\includegraphics[width=1.0\textwidth]{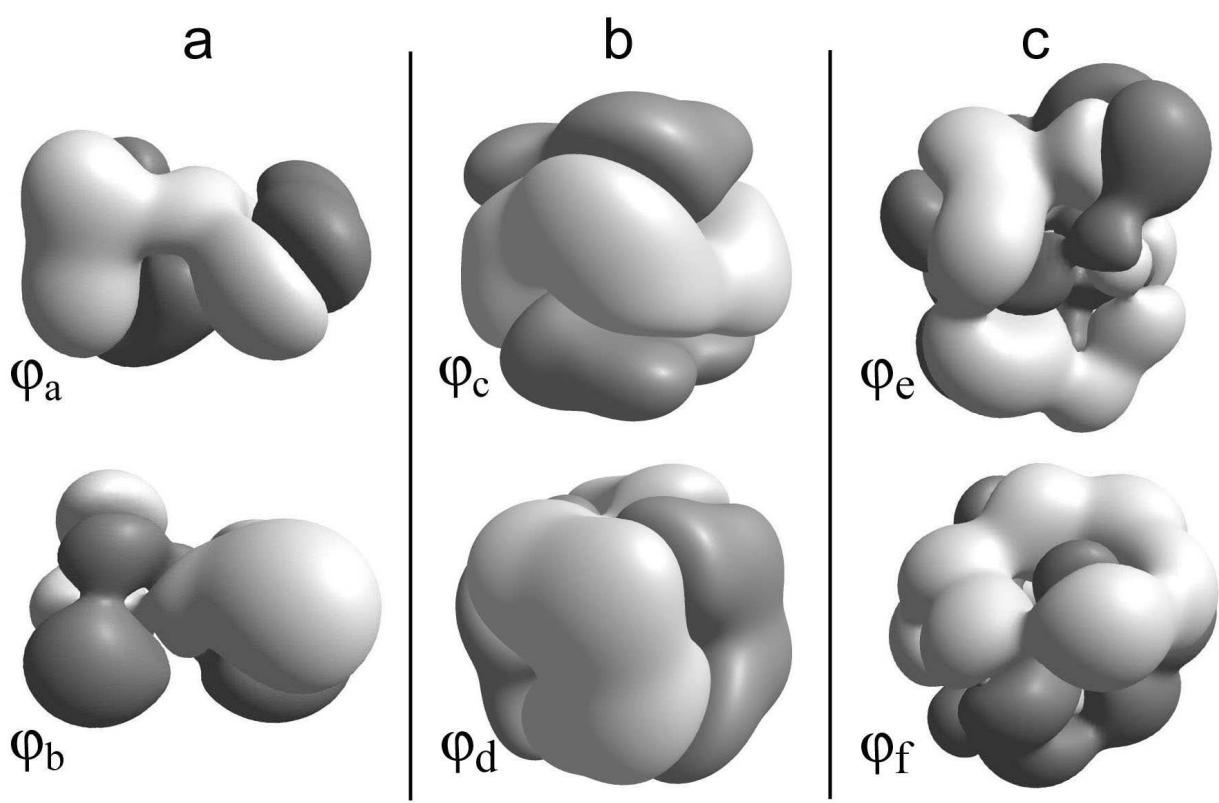}
\caption{Some valence $\pi$ orbitals of cage-like (H$_2$0)$_n$ clusters composed of fused or joined molecular rings and involving tri- and tetracoordinated water molecules: (a) $n$ = 5, (b) $n$ = 8, and (c) $n$ = 12.} \label{fig10}
\end{figure}

However, the situation dramatically changes when one goes to cage-like clusters. HF molecules do not agglomerate to form stable three-dimensional cages. By contrast, water molecules form three-dimensional networks. Mean H-bond energies in such systems are close to or even higher than 6 kcal/mol, and the bond energies per molecule are close to 8.5 kcal/mol., i.e., of the same order as in the most stable folded (HF)$_n$ clusters. The reason becomes clear if we consider Fig. \ref{fig10} where orbitals of (H$_2$O)$_5$ cluster, which involves a tetracoordinated molecule, and those of (H$_2$O)$_8$ and (H$_2$O)$_{12}$ cages composed of fused molecular rings, are shown. One of the orbitals of (H$_2$O)$_5$ ($\varphi_a$ in Fig. \ref{fig10}a) looks like two concatenated H letters, one leg of each being normal to the residual T-shaped segment. Such orbital provides $\pi$-electron binding in the domains of all six H-bonds. Two other orbitals (nearly degenerate) look like concatenated twisted Z-shaped patterns (Fig. \ref{fig10}a shows one of them, $\varphi_b$) and provide $\pi$ binding within one structural triangle and between two residual molecules. Similarly to the above orbitals of ring-like clusters, these orbitals have generally $\pi$ character and a mixed $\pi$/$\sigma$ character in the H-bond domains. Such flexibility of orbitals is determined by the contributions of $\phi_3^w$, $\phi_4^w$, and $\phi_5^w$ functions with different weights. Thus, a relatively small cluster that involves a tetracoordinated joint molecule is stabilized by $\pi$ binding between all molecules. Recall that, in the case of ammonia molecules, such binding was impossible.

Orbitals that span the whole or large part of the structure can be found also in (H$_2$O)$_8$ and (H$_2$O)$_{12}$ cages. Figure \ref{fig10} (b, c) shows four such orbitals. In octamer, there are orbitals of diverse shapes, and here, the principally new situation is observed. First of all, there is an orbital ($\varphi_c$), which looks like a sandwich and describes conjugated $\pi$-binding systems of two four-membered rings that form two opposite faces of the cube (let us conditionally refer to them as the bottom and top). Another orbital of the same kind ($\varphi_d$) describes binding within four tetramolecular rings that form side faces of the cube. There are also orbitals that describe $\pi$ binding in the domains of dimeric edges of the cube. Note that in all the cases, the orbitals have bonding or antibonding $\sigma$ character in the directions normal to $\pi$ bonds. This is the pronounced manifestation of the aforementioned flexibility of the electron density distribution in water clusters. 

In the case of (H$_2$O)$_{12}$, let us consider also only two orbitals ($\varphi_e$ and $\varphi_f$). At first glance, they are very similar to those of plain rings. In fact, they are. In one of them, interlaced constant-sign domains are composed of six- and five-membered rings connected via a typical $\pi$ bonding segment of a tetramolecular ring. In the other orbital, two concatenated constant-sign patterns are based on a six-membered ring fused with five- and four-membered rings on opposite edges. As a result, there appears a conjugated $\pi$ system that spans four structural rings, which in turn form a closed or open spatial hoop. Some other cluster orbitals look like interlaced or concatenated variously bent segments, and the corresponding electrons provide binding within four- to eight-molecular subsystems of the cluster. 

These results explain why cage-like structures are more stable than rings. As soon as the number of molecules is sufficient for the formation of a cage composed of fused rings whose (mean) molecular planes are either nearly orthogonal to each other (as in small (H$_2$O)$_n$ clusters with $n$ = 6 or 8) or form angles up to ca. 135$^{\circ}$ (as in (H$_2$O)$_n$ clusters with $n$ = 12 or 20), such arrangement becomes preferable. A large number of cluster orbitals have mixed nature: they act as $\sigma$ bonding (or antibonding) orbitals in one plane (ring) and as $\pi$ bonding orbitals in the orthogonal subspace. It is worth noting that in this orthogonal subspace, a continuous delocalized system, which spans several structural rings, is formed. The typical range of interplanar angles is determined by the mutual orientation of the initial molecular orbitals and their consistent contributions to $\sigma$ and $\pi$ binding. Moreover, even those orbitals, which look like antibonding in O--H\dots O region, have such spatial character that act as bonding in O--H and H\dots O subdomains. 

Thus, even in cage-like clusters, there are delocalized conjugated $\pi$ systems, which span either individual molecular rings or hoops composed of fused rings. (Recall that in the case of hydrogen fluoride, a sequence of quasi-aromatic segments rather than a continuous system can only be formed.)  Besides that, tetracoordinated water molecules can be effective joints between molecular rings in a three-dimensional network. Hence, diverse structural groups can be present in water ensembles as long-living structure units. The strength and lifetime of an individual H-bond depend on its involvement in the conjugated $\pi$ system of a larger spatial fragment. Though it is a time consuming task, it is quite possible to determine all long-living structure fragments. However, it is already clear that fused molecular rings involved in closed or open hoops should have the largest lifetimes among long-living species due to general delocalized $\pi$ binding.

\subsection{Ionic clusters}

Character of the electron density redistribution that accompanies the formation of (HF)$_n$, (H$_2$O)$_n$, and (NH$_3$)$_n$ clusters is generally similar, and peculiarities are determined by the degeneracy of molecular orbitals and the {\textquotedblleft}part{\textquotedblright} of the electron density that is not involved in intramolecular binding. Then, it is interesting to consider ionic clusters, and primarily HF$_2^-$ and H$_5$O$_2^+$, which are known to be stabilized by the strongest hydrogen bonds.

\begin{figure}[ht] \center
\includegraphics[width=0.65\textwidth]{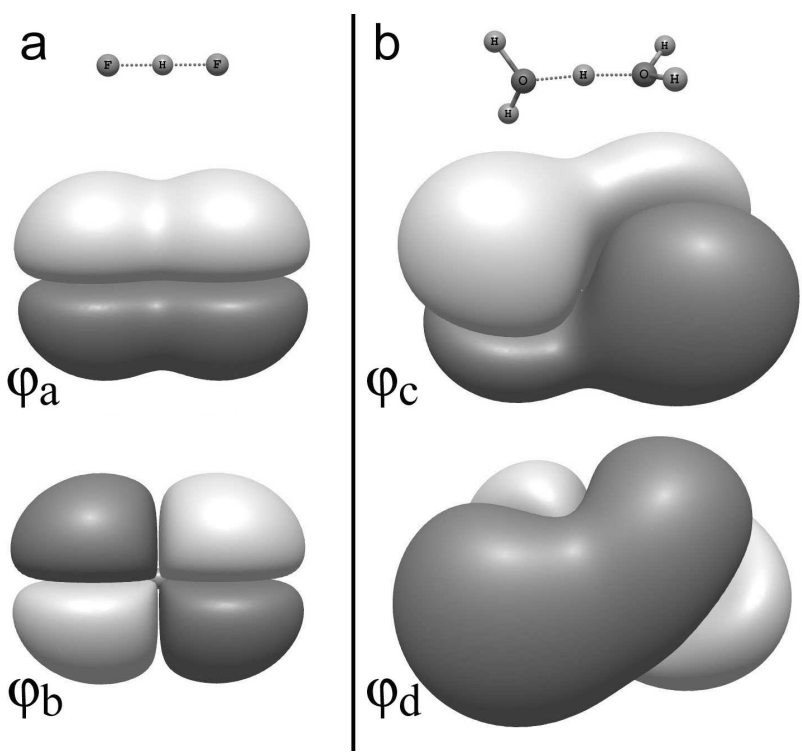}
\caption{Typical $\pi$ orbitals of (a) HF$_2^-$ and (b) H$_5$O$_2^+$.} \label{fig11}
\end{figure}

In HF$_2^-$, the binding is provided by electrons described by a $\sigma$-orbital and two degenerate mutually orthogonal $\pi$ orbitals (Fig. \ref{fig11}a shows one of them, $\varphi_a$). Moreover, zeros of two corresponding antibonding $\pi$ orbitals accurately coincide with the proton position (Fig. \ref{fig11}a shows one of them, $\varphi_b$). Recall that in neutral molecular clusters, an  alternation-sign surface could pass near a proton, but did not involve it. As a result, judging from the shape of the orbitals, one can say that there is some kind of a triple bond between two fluorine atoms. Moreover, both bonding and antibonding orbitals describe electrons that provide binding in F\dots H domains. 

The quasi-triple F\dots F bond and the nearly zero electron density of antibonding orbitals in the vicinity of the bridge proton represent the limiting variant of the effects, which are less pronounced in neutral clusters. However, it is the effects that clearly explain the experimental NMR signatures of hydrogen bonds, namely, proton deshielding and spin-spin coupling of the nuclei of electronegative atoms.

A similar picture (though more complex) is observed in the case of H$_5$O$_2^+$. In this ion, there are five cluster orbitals, each of which look like two embraced constant-sign segments (two of them are shown in Fig. \ref{fig11}b, $\varphi_c$ and $\varphi_d$). Accordingly, each orbital describes electronic binding of two H$_2$O fragments (with the zeros close to the position of the central proton). Thus, electrons of all five orbitals contribute to the binding in the system. As a result, the bonds in H$_5$O$_2^+$ ion are energetically very close to those in HF$_2^-$ ion. 

\begin{figure}[ht] \center
\includegraphics[width=0.8\textwidth]{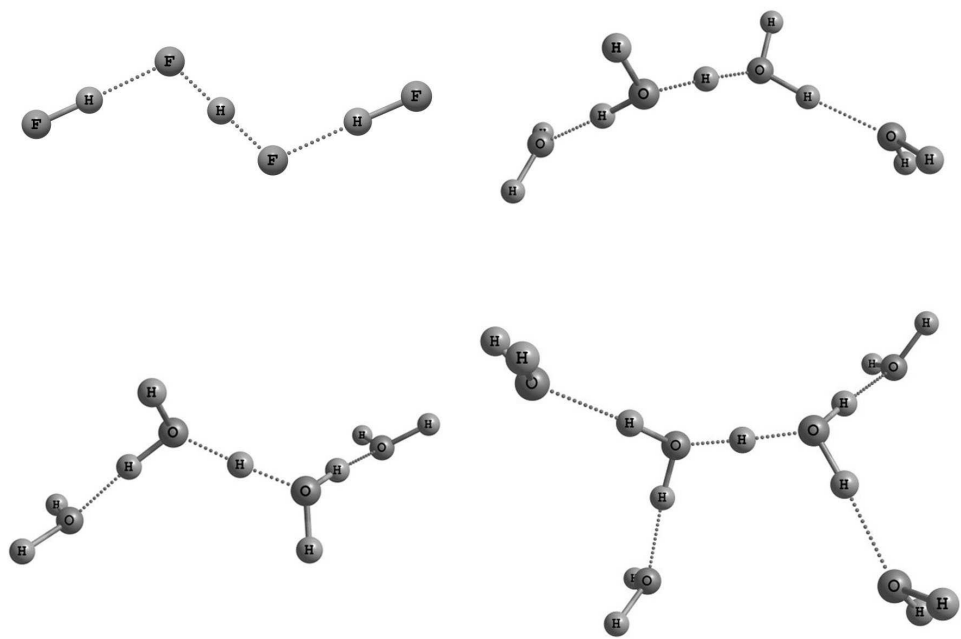}
\caption{Examples of F$^-$(HF)$_n$ and H$^+$(H$_2$O)$_n$ structures that involve stable HF$_2$ and H$_5$O$_2$ fragments respectively.} \label{fig12}
\end{figure}

Being highly sensitive to the arrangement of nuclei, the peculiar electron density distribution in these ionic structures survives only in the absence of external perturbation or when the perturbation has the same symmetry (no less than $\mathbb{C}_2$). Some examples of clusters, in which HF$_2$ or H$_5$O$_2$ fragment steadily exists, are shown in Fig. \ref{fig12} (there is much less freedom in constructing HF-based clusters). Note that in all systems, there is still a quasi-triple bond between two central oxygen or fluorine nuclei. However, for the spatial reasons, this kind of binding can by no means be spread on the residual structure. Therefore, all other hydrogen bonds are characterized by nearly the same electron density distribution as in neutral clusters.

Thus, hydrogen atom plays the role of a bridge between two neighboring electronegative atoms, i.e., acts as a binding joint between their electron density {\textquotedblleft}clouds{\textquotedblright}. This role is more pronounced in the ionic systems, but could already be deduced from the electron density distributions in neutral clusters. If we consider a general situation, the formation of an X--H\dots Y--Z hydrogen bond is determined by the overlap of X--H and Y--Z bonding orbitals (or Y lone pair orbital); and hydrogen atom acts as a bridge.

\section{Conclusions}

The discovered peculiarities of the electron density distribution in the hydrogen bond domains enable us to analyze some criteria and characteristics of the bonds. 

First of all, it is reasonable to note that there is always $\sigma$ and $\pi$ binding in H-bond domains. \textit{Hydrogen bonds are usually quasi-double and sometimes (as in the case of HF$_2^-$ and H$_5$O$_2^+$ ions) quasi-triple}. 

H-bonds should be divided in \textit{two classes}, namely, (i) separate bonds formed between two certain particles and (ii) bond networks that unite a large number of molecules and are stable when a delocalized conjugated $\pi$ system appears. Networks can be formed by hydrogen fluoride, water, and ammonia molecules, as well as any molecules that involve sterically accessible O--H or N--H functional groups. The most stable building blocks of the networks are those with \textit{delocalized quasi-aromatic $\pi$-systems}. 

Insofar as the \textit{proton in X--H\dots Y fragment acts as a bridge} between the electron density {\textquotedblleft}clouds{\textquotedblright} of X and Y, and the quasi-double or quasi-triple bond is formed between X and Y, it is natural that the bridge nuclei should lie on the X\dots Y line. Then, the closer the X--H\dots Y angle to 180$^{\circ}$, the stronger the bond. At the same time, the interaction is the strongest when X--H bond in the donor molecule and the orbital in the acceptor molecule that provides the main contribution to $\sigma$ binding have a common symmetry axis. This means that there is always a preferable orientation of the proton-acceptor molecule with respect to the H-bond direction, and the known  directionality of hydrogen bonds becomes a natural consequence. 

The discovered peculiarities of the electron density distribution in H-bond domains clarify the origin of the NMR signatures of the bonds, particularly deshielding of H-bond protons and spin–spin interaction between X and Y. As we could see, electrons described by $\sigma$ and $\pi$ orbitals actually bind X and Y, especially when a conjugated system forms. At the same time, in many cases, zeros of both $\sigma$  and $\pi$ functions are close to the position of the bridge proton and coincide with it in HF$_2^-$ and H$_5$O$_2^+$ ions. 

\textit{The existence of conjugated $\pi$ system in molecular clusters determines the cooperativity phenomenon. The spatial extension of cooperative effects depends on the extension of the delocalized conjugated $\pi$ system, and the effects are most pronounced within the regions of quasi-aromatic $\pi$-binding typical of structural rings and closed or open hoops composed of fused rings.}

\end{document}